\documentclass{svmult}

\usepackage{makeidx}     
\usepackage{graphicx}    
\usepackage{multicol}    

\makeindex             

\usepackage[sort,numbers]{natbib}
\usepackage{amsthm}
\usepackage{amsmath}
\usepackage{amssymb}
\theoremstyle{plain}
\newtheorem{thm}{Theorem}
  \theoremstyle{definition}
  \newtheorem{defn}[thm]{Definition}
  \theoremstyle{remark}
  \newtheorem{rem}[thm]{Remark}
  \theoremstyle{plain}
  \newtheorem{lem}[thm]{Lemma}
  \theoremstyle{plain}
  \newtheorem{cor}[thm]{Corollary}
\usepackage{units}

\def\bY{\mathbb{Y}}
\def\bX{\mathbb{X}}

\begin{document}

\title*{Combinatorial information distance}
\author{Joel Ratsaby}
\institute{\it Department of Electrical and Electronics Engineering, Ariel University Center\\
Ariel 40700, ISRAEL
\\
\texttt{ratsaby@ariel.ac.il}}

%
%
\maketitle

\abstract{
Let $\left|A\right|$ denote the cardinality of a finite set $A$.
For any real number $x$ define $t(x)=x$ if $x\geq1$ and $1$ otherwise.
For any finite sets $A,B$ let $\delta(A,B)$ $=$ $\log_{2}\left(t\left(\left|B\cap\overline{A}\right|\left|A\right|\right)\right)$.
We define\footnote{This appears as Technical Report \#	arXiv:0905.2386v4. A shorter version appears in the {Proc. of Mini-Conference on Applied Theoretical Computer Science (MATCOS-10)}, Slovenia, Oct. 13-14, 2010.  } a new cobinatorial distance $d(A,B)$ $=$ $\max\left\{ \delta\left(A,B\right),\delta\left(B,A\right)\right\} $
which may be applied to measure the distance between binary strings
of different lengths. The distance is based on a classical combinatorial notion of information introduced by Kolmogorov.
}

\section{\label{Sec1}Introduction}

A basic problem in pattern recognition \cite{Duda-2000-PCL} is to
find a numerical value that represents the dissimilarity or `distance'
between any two input patterns of the domain. For instance, between
two binary sequences that represent document files or between genetic
sequences of two living organisms. There are many distances defined
in different fields of mathematics, engineering  and computer and information sciences  \cite{Deza-09}. A good distance is one which picks out only the `true'
dissimilarities and ignores those that arise from irrelevant attributes
or due to noise. In most applications the design of a good distance
requires inside information about the domain, for instance, in the
field of information retrieval \cite{YAtes99} the distance between
two documents is weighted largely by words that appear less frequently
since the words which appear more frequently are less informative.
The ubiquitous Levenshtein-distance \cite{Levenshtein1966} measures
the distance between two sequences (strings) as the minimal number
of edits (insertion, deletion or substitution of a single character)
needed to transform one string into another. Approximate string matching \cite{Nava} is an area that uses such edit-distances to find
matches for short strings inside long texts. Typically, different
domains require the design of different distance functions which take
such specific prior knowledge into account. It can therefore be an
expensive process to acquire expertise in order to formulate a good
distance. The paper of \cite{LZ76} introduced a notion of complexity
of finite binary string which does not require any prior knowledge
about the domain or context represented by the string (this is sometimes
referred to as the \emph{universal} property). This complexity (called
the production complexity of a string) is defined as the minimal number
of copy-operations needed to produce the string from a starting short-string
called the base. This definition of complexity is related to Levenshtein-distance
mentioned above. It is proportional to the number of distinct phrases
and the rate of their occurrence along the sequence. There has been
some work on using the LZ-complexity to define a sequence-distance
measure in bioinformatics \cite{Sayood2003}. Other applications of
the LZ-complexity include: approximate matching of strings \cite{Nava},
analysis of complexity of biomedical signals \cite{Aboy2006}, recognition
of structural regularities \cite{Orlov2004}, characterization of
DNA sequences \cite{Gusev1999} and responses of neurons to different
stimuli \cite{Amigo2003}, study of brain function \cite{Wu-1991}
and brain information transmission \cite{Xu-97} and EEG complexity
in patients \cite{Abasolo-2006}.

In the current paper we introduce a distance function between two
strings which also possesses this universal property. Our approach
is to consider a binary string as a \emph{set} of substrings \cite{Ratsaby-CRI-2008}.
To represent the complexity of such a set we use the notion of combinatorial
entropy \cite{Ratsaby2006a} and introduce a new set distance function.
We proceed to describe some fundamental concepts concerning entropy
and information of sets.

\section{Entropy and information of a set\label{sec:Entropy-and-information}}
\label{sec2}

Kolmogorov \cite{Kolmogorov65} investigated a non-stochastic measure
of information for an object $y$. Here $y$ is taken to be any element
in a finite space $\bY$ of objects. He defines the `entropy' of $\bY$
as $H(\bY)=\log\left | \mathbb{Y}\right|$ where $\left| \mathbb{Y}\right | $ denotes the cardinality
of $\bY$ and all logarithms henceforth are taken with respect to
$2$. 

As he writes, if it is known that $\bY=\{y\}$ then this provides
$\log\left| \mathbb{Y}\right | $ bits of `information' or in his words {}``this
much entropy is eliminated''. To represent partial information about
$\bY$ based on another information source $\bX$ let $R=\bX\times\bY$
be a general finite domain and consider a set \begin{equation}
A\subseteq R\label{A}\end{equation}
that consists of all permissible pairs $(x,y)\in R$ (in the usual
probabilistic-based representation of information this is analogous
to having a uniform prior probability distribution over a certain
region of the domain). The entropy of $\bY$ is defined as \[
H(\mathbb{Y})=\log\left|\Pi_{\mathbb{Y}}(A) \right|\]
 where $\Pi_{\mathbb{Y}}(A)\equiv\{y\in\mathbb{Y}:(x,y)\in A\text{ for some }x\in\mathbb{X}\}$
denotes the projection of $A$ on $\bY$. Consider the restriction
of $A$ on $\bY$ based on $x$ which is defined as \begin{equation}
Y_{x}=\{y\in\bY:(x,y)\in A\},\; x\in\Pi_{\bX}(A)\label{Ax}\end{equation}
then the conditional combinatorial entropy of $\bY$ given $x$ is
defined as \begin{equation}
H(\mathbb{Y}|x)=\log\left| Y_{x} \right| .\label{Hxy}\end{equation}
Kolmogorov defines the information conveyed by $x$ about $\bY$ by
the quantity \begin{equation}
I(x:\bY)=H(\bY)-H(\bY|x).\label{KIxy}\end{equation}
In \cite{Ratsaby_IW} an alternative view of $I(x:\bY)$ is defined
as the information that a set $Y_{x}$ conveys about another set $\bY$
satisfying $Y_{x}\subseteq\bY$. Here the domain $R$ is defined based
on the previous set $A$ as $R=\Pi_{\bY}(A)\times\Pi_{\bY}(A)$ which
consists of all permissible pairs $(y,y')$ of objects. Knowledge
of $x\in\bX$ means knowing the set $A_{x}\subseteq R$, $A_{x}=\{(y,y'):y\in\Pi_{\bY}(A),y'\in Y_{x}\}$.
The information between $Y_{x}$ and $\bY$ is then defined as \begin{eqnarray}
I(Y_{x}:\bY) & = & \log\left(\left|  \Pi_{\bY}(A)\right| ^{2}\right)-\log\left|  A_{x} \right | \nonumber \\
 & = & \log\left(\left| \Pi_{\bY}(A)\right | ^{2}\right)-\log(\left| \Pi_{\bY}(A)\right | \left| Y_{x} \right| ).\label{eq:btts}\end{eqnarray}
Clearly, $I(Y_{x}:\bY)=I(x:\bY)$. Note that $I(Y_{x}:\bY)$ measures
the difference in description length of any \emph{pair} of objects
$\left(y,y'\right)\in\Pi_{\bY}(A)\times\Pi_{\bY}(A)$ when no 'labeling'
information exists versus that when there exists information which
labels one of them as being an element of $Y_{x}$. Thus the second
term in (\ref{eq:btts}) can be viewed as the conditional combinatorial
entropy of $\Pi_{\bY}(A)$ given the set $Y_{x}$. In \cite{Ratsaby2006a,Ratsaby_IW,RATSABY_DBLP:conf/sofsem/Ratsaby07}
this is used to extend Kolmogorov's combinatorial information to a
more general setting where knowledge of $x$ still leaves some vagueness
about the possible value of $y$. 

While the distance that we introduce in this paper is general enough
for any objects, our interest is to introduce a combinatorial distance
for binary strings. We henceforth drop the finiteness constraint on $\mathbb{X}$ and $\mathbb{Y}$  and refer to $\bX=\left\{ 0,1\right\} ^{*}$
as the set of finite binary strings $x$. Each string $x\in\bX$ is a \emph{description}
of a corresponding set $Y_{x}$ contained in  the set $\bY$ of objects $y$.
Our approach to defining a distance between two binary strings $x$
and $x'$ is to relate them to sets of objects and then measure the
distance between the two corresponding sets.  Denote by $\mathcal{P}_F(X)$ the set of all finite subsets of a set $X$.
Let  $M:\mathbb{X}\rightarrow\mathcal{P}_F(\mathbb{Y})$ be
a function which defines how a description (binary string) $x$ yields a set $Y_{x}\subseteq\bY$.
In general, $M$ may be a many-to-one function since there may be
several strings (viewed as descriptions of the set) of different lengths
for a given set. In the context of the above, we now consider a permissible
pair $(x,y)\in A$ to be one which consists of an object $y$ that
is contained in a set $Y_{x}$ which is described by $x$. Clearly,
not every possible pair $(x,y)$ is permissible, as for instance,
if $y'\not\in Y_{x}$ then $(x,y')$ is not permissible.

In the next section we introduce a combinatorial information distance.
We start with a distance for general sets and then apply it as a distance
between binary strings.

\section{The distance}

In what follows, $\Omega$ is a given non-empty set which serves as
the domain of interest. The cardinality of any set $A$ is denoted
by $\left|A\right|$ and the set of all finite subsets of $\Omega$
is denoted by $\mathcal{P}_{F}(\Omega)$. Define $t:\mathbb{R}\rightarrow\mathbb{R}$
as follows: \[
t(x)=\left\{ \begin{array}{cc}
x & \text{if} ~  x\geq1\\
1 &   \text{otherwise .}\end{array}\right.\]

\begin{defn}
For each pair of finite sets $A,B\subset\Omega$ define the following
function $\delta:\mathcal{P}_{F}(\Omega)\times\mathcal{P}_{F}(\Omega)\rightarrow\mathbb{N}_{0}$
which maps a pair of finite sets into the non-negative integers, \[
\delta(A,B):=\log\left(t\left(\left|B\cap\overline{A}\right|\left|A\right|\right)\right)\]
where $\overline{A}$ denotes the complement of the set $A$ and $\log$
is with respect to base $2$. It is simple to realize that $\delta(A,B)$
equals $\log\left(\left|B\cap\overline{A}\right|\left|A\right|\right)$
with the exception when $A$ or $B$ is empty or $B\subseteq A$. \end{defn}
\begin{rem}
\label{rem:Note-that-}Note that $\delta$ is non-symmetric, i.e.,
$\delta(A,B)$ is not necessarily equal to $\delta(B,A)$. Also, $\delta(A,B)=0$
when $B\subseteq A$ (not only when $A=B$).
\end{rem}
From an information theoretical perspective \cite{Kolmogorov65} the
value $\log\left|B\cap\overline{A}\right|$ represents the additional
description length (in bits) of an element in $B$ given \emph{a priori
}knowledge of the set $A$. Hence we may view $A$ as a partial 'dictionary'
while the part of $B$ that is not included in $A$ takes an additional
$\log\left|B\cap\overline{A}\right|$ bits of description given $A$.

The following set will serve as the underlying space on which we will
consider our distance function. It is defined as

\begin{eqnarray*}
\mathcal{P}_{F}^{+}(\Omega) & := & \mathcal{P}_{F}(\Omega)\setminus\left\{ A\subset\Omega:\left|A\right|\leq1\right\} .\end{eqnarray*}
It is the power set of $\Omega$ but without the empty set and singletons.
We note that in practice for most domains, as for instance the domain
of binary strings considered later, the restriction to sets of size
greater than $1$ is minor. 

The following lemma will be useful in the proof of Theorem \ref{main-theom}.
\begin{lem}
\label{lem:The-function-lem2} The function $\delta$ satisfies the
triangle inequality on any three elements $A$, $B$, $C\in\mathcal{P}_{F}^{+}(\Omega)$
none of which is strictly contained in any of the other two.\end{lem}
\begin{proof}
Suppose $A,B,C$ are any elements of $\mathcal{P}_{F}^{+}(\Omega)$
satisfying the given condition. It suffices to show that \begin{eqnarray}
\delta(A,C) & \leq & \delta(A,B)+\delta(B,C).\label{eq:main}\end{eqnarray}
First we consider the special case where the triplet has an identical
pair. If $A=C$ then by Remark \ref{rem:Note-that-} it follows that
$\delta(A,C)=0$ which is a trivial lower bound so (\ref{eq:main})
holds. If $A=B$ then $\delta(A,B)=0$ and both sides of (\ref{eq:main})
are equal hence the inequality holds (similarly for the case of $B=C$). 

Next we consider the case where each of the following three quantities
satisfies \begin{eqnarray}
\#\left(C\cap\overline{A}\right),\;\#\left(B\cap\overline{A}\right),\;\#\left(C\cap\overline{B}\right) & \geq & 1.\label{eq:cb}\end{eqnarray}
By definition of $\mathcal{P}_{F}^{+}(\Omega)$ we have $\left|A\right|\geq2$
hence \[
\delta(A,C)=\log\left(t\left(\left|C\cap\overline{A}\right|\left|A\right|\right)\right)=\log\left(\left|C\cap\overline{A}\right|\left|A\right|\right)=\log\left|C\cap\overline{A}\right|+\log\left|A\right|.\]
Next, we claim that $C\cap\overline{A}\subseteq\left(B\cap\overline{A}\right)\cup\left(C\cap\overline{B}\right)$.
If $x\in C\cap\overline{A}$ then $x\in C$ and $x\in\overline{A}$.
Now, either $x\in B$ or $x\in\overline{B}$ . If $x\in B$ then because
$x\in\overline{A}$ it follows that $x\in B\cap\overline{A}$. If
$x\in\overline{B}$ then because $x\in C$ it follows that $x\in C\cap\overline{B}$.
This proves the claim. Next, we have \begin{eqnarray*}
\delta(A,B)+\delta(B,C) & = & \log\left|A\right|+\log\left|B\cap\overline{A}\right|+\log\left|B\right|+\log\left|C\cap\overline{B}\right|.\end{eqnarray*}
It suffices to show that \begin{equation}
\log\left|C\cap\overline{A}\right|\leq\log\left|B\cap\overline{A}\right|+\log\left|C\cap\overline{B}\right|+\log\left|B\right|.\label{eq:ca}\end{equation}
We claim that if three non-empty sets $X,Y,Z$ satisfy $X\subseteq Y\cup Z$
then $\log\left|X\right|\leq\log\left(2\left|Y\right|\left|Z\right|\right)$.
To prove this, it suffices to show that $\left|X\right|\leq2\left|Y\right|\left|Z\right|$.
That this is true follows from $\left|X\right|\leq\left|Y\cup Z\right|\leq\left|Y\right|+\left|Z\right|$$\leq\left|Y\right|\left|Z\right|+\left|Z\right|\left|Y\right|=2\left|Y\right|\left|Z\right|$.
By (\ref{eq:cb}), we may let $X=C\cap\overline{A}$, $Y=B\cap\overline{A}$
and $Z=C\cap\overline{B}$ and from both of the claims it follows
that \begin{equation}
\left|C\cap\overline{A}\right|\leq2\left|B\cap\overline{A}\right|\left|C\cap\overline{B}\right|.\label{eq:logs}\end{equation}
Taking the log on both sides of (\ref{eq:logs}) and using the inequality
$2\leq\#B$ (which follows from $B\in\mathcal{P}_{F}^{+}(\Omega)$)
we obtain \[
\log\left|C\cap\overline{A}\right|\leq1+\log\left|B\cap\overline{A}\right|+\log\left|C\cap\overline{B}\right|\leq\log\left|B\right|+\log\left|B\cap\overline{A}\right|+\log\left|C\cap\overline{B}\right|.\]
This proves (\ref{eq:ca}).
\end{proof}
Next, we define the information set-distance.
\begin{defn}
\label{def:info-set-dist}For any two finite non-empty sets $A,B$
define the\emph{ information set-distance} as\[
d\left(A,B\right):=\max\left\{ \delta\left(A,B\right),\delta\left(B,A\right)\right\} .\]
In the following result we show that $d$ satisfies the properties
of a semi-metric.\end{defn}
\begin{thm}
\label{main-theom}The distance function $d$ is a semi-metric on
$\mathcal{P}_{F}^{+}(\Omega)$. It satisfies the triangle inequality
for any triplet $A,B,C\in\mathcal{P}_{F}^{+}(\Omega)$ such that no
element in the triplet is strictly contained in any of the other two. \end{thm}

\begin{proof}
That the function $d$ is symmetric is clear from its definition.
From Remark \ref{rem:Note-that-} it is clear that for $A=B$, $\delta(A,B)=\delta(B,A)=0$
hence $d(A,B)=0$. 
Consider any pair of sets $A,B\in\mathcal{P}_{F}^{+}(\Omega)$ such that $A\neq B$.
If $A\cap B=\emptyset$ or
$A\subset B$ or $B\subset A$ then at least one of  the two values
 $\delta(A,B)$ or  $\delta(B,A)$ is greater than zero so $d(A,B)>0$. 
This
means that $d$ is a semi-metric on $\mathcal{P}_{F}^{+}(\Omega)$.
Next, we show that it satisfies the triangle inequality for any triplet
$A,B,C\in\mathcal{P}_{F}^{+}(\Omega)$ such that no element is strictly
contained in any of the other two. For any non-negative numbers $a_{1}$,
$a_{2}$, $a_{3}$, $b_{1}$, $b_{2}$, $b_{3}$, that satisfy \begin{eqnarray}
a_{1} & \leq & a_{2}+a_{3}\nonumber \\
b_{1} & \leq & b_{1}+b_{2},\label{eq:abb}\end{eqnarray}
we have\begin{eqnarray*}
\max\left\{ a_{1},b_{1}\right\}  & \leq & \max\left\{ a_{2}+a_{3},b_{2}+b_{3}\right\} \\
 & \leq & \max\big\{ \max\left\{ a_{2},b_{2}\right\} +\max\left\{ a_{3},b_{3}\right\} ,\\
 && \max\left\{ b_{2},a_{2}\right\} +\max\left\{ b_{3},a_{3}\right\} \big\} \\
 & = & \max\left\{ a_{2},b_{2}\right\} +\max\left\{ a_{3},b_{3}\right\} .\end{eqnarray*}
From Lemma \ref{rem:Note-that-} it follows that (\ref{eq:abb}) holds
for the following: $a_{1}=\delta(A,C)$, $b_{1}=\delta(C,A)$, $a_{2}=\delta(A,B)$,
$b_{2}=\delta(B,A)$, $a_{3}=\delta(B,C)$, $b_{3}=\delta(C,B)$.
This yields \[
d(A,C)\leq d(A,B)+d(B,C)\]
hence $d$ satisfies the triangle inequality for such a triplet.\end{proof}
\begin{rem}
Currently, it is an open question as to whether a normalized version of the distance  $d$
exists such that the properties stated in Theorem \ref{main-theom} are still satisfied.
\end{rem}

\section{Distance between strings}

Let us now define the distance between two binary strings. In this
section, we take $\Omega$ to be a set $\mathbb{Y}$ of objects. Denote
by $\bX$ the set of all (finite) binary strings. Our approach to
defining a distance between two binary strings $x$, $x'\in\mathbb{X}$
is to relate them to subsets $Y_{x},Y_{x'}\in\mathcal{P}_{F}^{+}(\mathbb{Y})$
and measure the distance between the two corresponding subsets. Each
string $x\in\bX$ is a \emph{description} of a corresponding set $Y_{x}\in\mathcal{P}_{F}^{+}(\Omega)$.
Define a function $M:\mathbb{X}\rightarrow\mathcal{P}_{F}^{+}(\mathbb{Y})$
which dictates how a string $x$ yields a set $M(x):=Y_{x}\subseteq\bY$.
In general, $M$ may be a many-to-one function since there may be
several strings (viewed as descriptions of the set) of different lengths
for a given set. 
\begin{defn}
\label{def:comb-dist}Let $\bX\times\bY$ be all possible string-object
pairs $(x,y)$ and let $M$ be any function $M:\mathbb{X}\rightarrow\mathcal{P}_{F}^{+}(\mathbb{Y})$.
If $x,x'\in\bX$ are two binary strings then the \emph{information}
\emph{set-distance} between them is defined as\[
d_{M}(x,x'):=d(M(x),M(x'))\]
where the function $d$ is defined in Definition \ref{def:info-set-dist}.
\end{defn}
The next result follows directly from Theorem \ref{main-theom}.
\begin{cor}
\label{main-cor}Let $\bY$ be a set of objects $y$ and $\bX$ a
set of all finite binary strings $x$. Let $M:\mathbb{X}\rightarrow\mathcal{P}_{F}^{+}(\mathbb{Y})$
be any function that defines the set $Y_{x}\subseteq\bY$ of cardinality
at least $2$ described by $x$, for all $x\in\mathbb{X}$. The information
set-distance $d_{M}(x,x')$ is a semi-metric on $\bX$ and satisfies
the triangle inequality for triplets $x$, $x'$,$x''$ whose sets
$M(x)$, $M(x')$, $M(x'')$ are not strictly contained in any of
the other two.
\end{cor}
As an example, consider a mapping $M$ that takes binary strings to
sets $Y$ in $\bY=\left\{ 0,1\right\} ^{k}$ (the $k$-cube) for some
fixed finite $k$. Denote by $k$-word a vertex on the cube. Consider
the following scheme for associating finite strings $x$ with sets:
given a string $x$, break it into non-overlapping $k$-words while,
if necessary, appending zeros to complete the last $k$-word. Let
the set $M(x)=Y_{x}$ be the collection of these $k$-words. For instance,
if $x=100100110$ then with $k=4$ we we obtain the set $Y_{x}=\left\{ 1001,0011,0000\right\} $.
If a string has $N>1$ repetitions of some $k$-word then clearly
only a single copy will be in $Y_{x}$. In this respect, $M$ eliminates
redundancy in a way that is similar to the method of \cite{LZ76}
which gives the minimal number of copy operations needed to reproduce
a string from a set of its substrings.

Another mapping $M$ may be defined by scanning a fixed window of
length $k$ across the string $x$ and collecting each substring (captured
in the window) as an element of the generated set $Y_{x}$. For instance,
suppose an alphabet has $26$ letters and there are $26^{n}$ possible
$n$-grams (substrings made of $n$ consecutive letters). If $x$
is a document then it can be broken into a \emph{set} $M(x)$ of $n$-grams.
Each letter is represented by $7$ bits. We extract words of length
$k=7n$ bits, starting with the first word in the string then  moving  $7$ bits to the right and extracting
 the next $k$-bit word, repetitively, until all words are collected. Thus
$d_{M}$ measures the distance between two documents. In comparison,
the $n$-gram model in the area of information retrieval \cite{YAtes99}
represents a document by a binary \emph{vector} of dimensionality
$26^{n}$ where the $i^{th}$ component is $1$ if the document contains
the $i^{th}$ particular $n$-gram and is $0$ otherwise. Here a similarity
(opposite of distance) between two documents is represented by the
inner product of their corresponding binary vectors. 

Yet another approach which does not need to choose a value for $k$
is to proceed along the line of work of \cite{LZ76}. Here we can
collect substrings of $x$ (of possibly different lengths) according
to a repetitive procedure in order to form the set $M(x)$ (in \cite{LZ76}
the cardinality of the set $M(x)$ is referred to as the complexity
of $x$).

Whichever scheme $M$ is used, to compute the information set-distance
$d_{M}(x,x')$ between two finite strings $x$ and $x'$ we first
determine the sets $M(x)$ and $M(x')$ and then evaluate their distance
according to Definition \ref{def:comb-dist} to be $d(M(x),M(x'))$.

\bibliographystyle{plain}


\end{document}